\documentclass[twocolumn,showpacs,preprintnumbers,superscriptaddress,amsmath,amssymb,nofootinbib]{revtex4}

\input psfig.sty

\usepackage{graphicx}
\usepackage{dcolumn}
\usepackage{bm}

\newcommand{\be}{\begin{equation}}
\newcommand{\ee}{\end{equation}}

\begin{document}

\title{A generalized equation of state for dark energy}

\author{E. M. Barboza Jr.} \email{edesio@on.br}

\author{J. S. Alcaniz} \email{alcaniz@on.br}

\affiliation{Observat\'orio Nacional, 20921-400, Rio de Janeiro -- RJ, Brasil}

\author{Zong-Hong Zhu} \email{zhuzh@bnu.edu.cn}

\affiliation{Department of Astronomy, Beijing Normal University, Beijing 100875, China}

\author{R. Silva} \email{raimundosilva@dfte.ufrn.br}


\affiliation{Departamento de F\'{\i}sica, Universidade Federal do Rio Grande do Norte, 59072-970 Natal - RN, Brasil}

\affiliation{Departamento de F\'{\i}sica, Universidade do Estado do Rio Grande do Norte, 59610-210, Mossor\'o - RN, Brasil}

\date{\today}

\begin{abstract}
A generalized parameterization $w_\beta(z)$ for the dark energy equation of state (EoS) is proposed and some of its cosmological consequences are investigated. We show that in the limit of the characteristic dimensionless parameter $\beta \rightarrow +1, 0$ and -1 some well-known EoS parameterizations are fully recovered whereas for other values of $\beta$ the proposed parameterization admits a wider and new range of cosmological solutions. We also discuss possible constraints on the $w_\beta(z)$ parameters from current observational data.

\end{abstract}


\maketitle

\section{Introduction}

The arrival of every new set of observational data has so far reconfirmed the current cosmic acceleration, which in turn poses to  cosmology a fundamental task of identifying and unveiling the cause of such a phenomenon. As is well known, in the context of Einstein's general relativity, this phenomenon is directly associated with the existence of new fields in high energy physics, the so-called dark energy or {quintessence} (see, e.g., \cite{rev} for recent reviews). 

Following this route, the dark energy equation of state (EoS), i.e., the ratio of its pressure to its energy density, $\omega(z) \equiv p/\rho$, has become one of the most searched numbers nowadays in theoretical and observational cosmology. This is so because if one could set (for some fundamental  principle or observational result) $\omega$ to be constant and exactly -1, then there would be a great probability of identifying the dark energy with the vacuum state of all existing fields in the Universe, i.e., the cosmological constant ($\Lambda$). Similarly, if a value $\omega(z) \neq -1$ is unambiguously found, then one could not only rule out $\Lambda$ but also seriously think of the dark pressure responsible for the cosmic acceleration as the potential energy density associated with a dynamical scalar field $\phi$~\footnote{The possibility $\omega(z) \neq -1$ still leads to two different routes, i.e., either the so-called \emph{quintessence} if $-1 < \omega(z) < -1/3$ \cite{quint} or  \emph{phantom} fields if $\omega(z) < -1$ \cite{phantom}. Both cases violate the strong energy condition $\rho + 3p > 0$, but the latter goes even further and also violates the null energy condition $\rho + p >0$ \cite{ec}.}.

In practice, at least three different approaches could be considered in order to find $\omega(z)$ from observations. The first and most direct one is to solve the scalar field equation for a particular theory. However, clearly such a procedure cannot provide a model-independent parameter space to be compared with the observational data. Another possibility is to build a functional form for $w(z)$ in terms of its current value $w_0$ and of its time-dependence $w' \equiv dw/d\ln a$ that avoids undesirable (and unphysical) behaviours in the past, present and future evolution~\cite{teg}. Recently, a number of EoS parameterizations have been discussed in the literature (see, e.g., \cite{teg,Huterer,astier,Weller,Efstathiou,Chevallier,Linder,goliah,otherP} and Refs. therein). As discussed in Ref. \cite{l1},  any functional form for $w(z)$ may in principle limit or even bias the physical interpretation of the data so that a third, parameter-free approach, such as binned EoS, decomposition into orthorgonal basis and principal component analysis, has also been considered as an alternative to EoS parameterization~\cite{l1}. Although promising, it is fair to say that such a procedure does not solve all the existing problems and may also introduce new ones, such as model dependence and uncertain signal-to-noise criteria (see \cite{l1} for more on this subject).

In this paper, we follow the second approach discussed above and consider a new parameterization for the dark energy EoS, which is characterized by a dimensionless parameter $\beta$. In the limits $\beta \rightarrow$ (-1, 0, +1), this new EoS form fully generalizes three of the most common EoS parameterizations investigated in the literature whereas $\forall$ $\beta \neq (-1, 0, +1)$ it admits a much wider range of solutions. Among these solutions, many of the different cosmological models that have been proposed to explain dark energy as well as new ones can be incorporated into the functional form here proposed. We emphasize that such flexibility and generality are particularly important to our research on $w(z)$ not only because they increase the range of possibilities to be tested but also because in principle they may reduce the possibility for misleading results an incorrect EoS parameterization can produce. 

This paper is organized as follows. In Secs. II and III we discuss some features of this new EoS parameterization and investigate its influence on the evolution of the dark energy--dark matter ratio and on the epoch of cosmic acceleration. In Sec. III we also test the viability of the parameterization proposed by investigating constraints on the 2-dimensional $w_0 - \beta$ and $w_\beta - \beta$ planes and on the 3-dimensional $w_0 - w_\beta - \beta$ space from distance measurements of type Ia supernovae (SNe Ia), measurements of the  baryonic acoustic oscillations (BAO) and cosmic microwave background (CMB), and measurements of the Hubble expansion $H(z)$ at low and intermediary redshifts. We end this paper by summarizing our main results in Sec. IV.

\section{New Parameterization}

Let us start by presenting some of the most investigated EoS parameterizations:
\begin{eqnarray}
\label{ps}
w(z) = \; \left\{
\begin{tabular}{l}
$w_0 + w_{\rm{P1}} z$ \quad  \quad  \quad  \quad   \hspace{0.17cm}(P1)  
 \hspace{0.35cm}\cite{Huterer,astier,Weller}\\
\\
$w_0+w_{\rm{P2}}\ln(1+z)$ \quad (P2)
\quad \cite{Efstathiou}\\
\\
$w_0+w_{\rm{P3}}z/(1+z)$ \quad (P3) \quad 
\cite{Chevallier, Linder}

\end{tabular}
\right.
\end{eqnarray}
where $w_0$ is the current value of the EoS parameter, and $w_{\rm{P}}$ (P = P1, P2, P3) are free parameters quantifying the time-dependence of the dark energy EoS, which must be adjusted by the observational data. Note that the EoS associated with $\Lambda$ can be always recovered by taking $w_{\rm{P}} = 0$ and  $w_0 = -1$ (see also \cite{otherP} for other parameterizations).

\begin{figure*}
\centerline{\psfig{figure=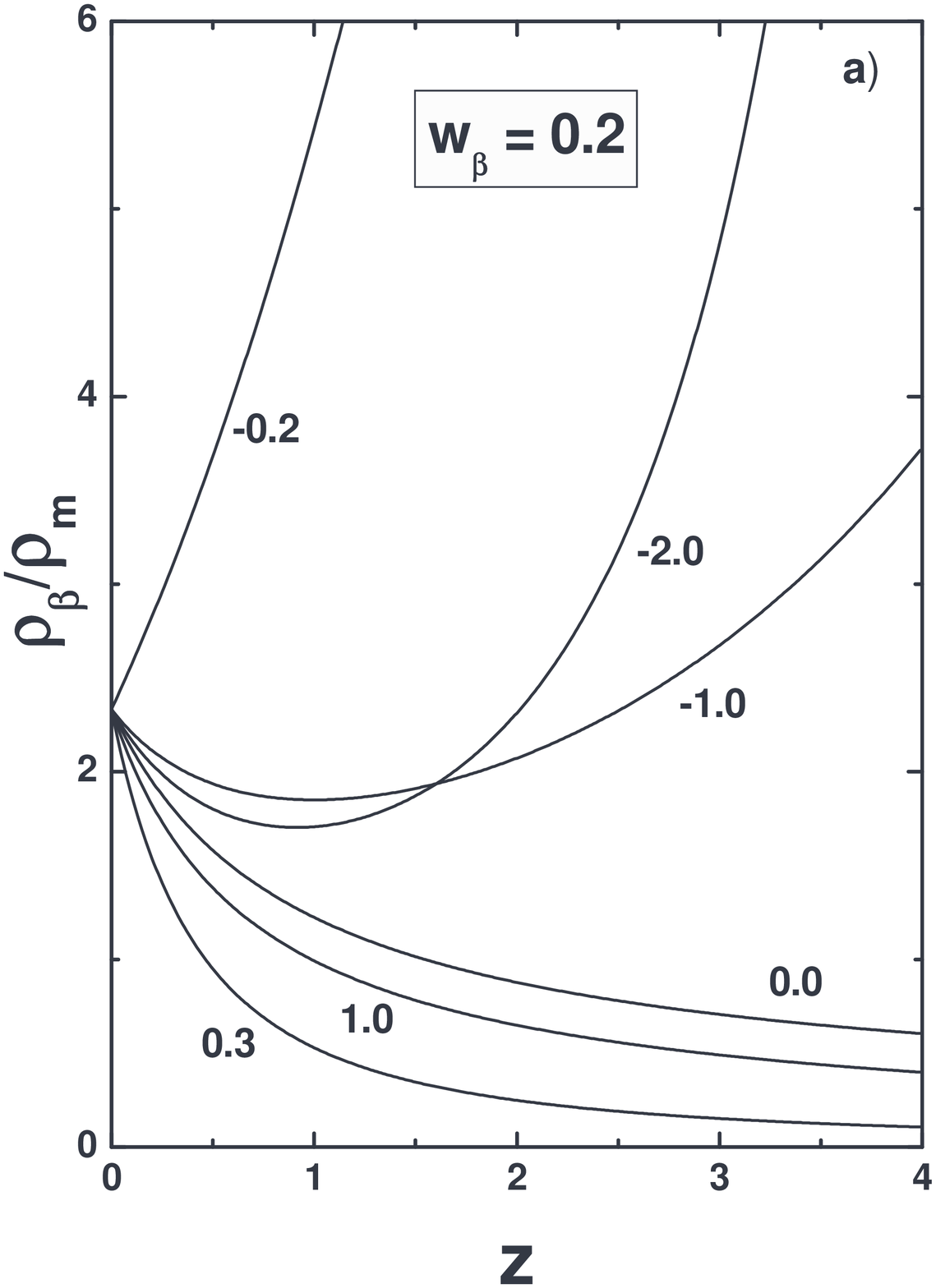,width=3.4truein,height=2.6truein,angle=0} 
\hspace{0.2cm}
\psfig{figure=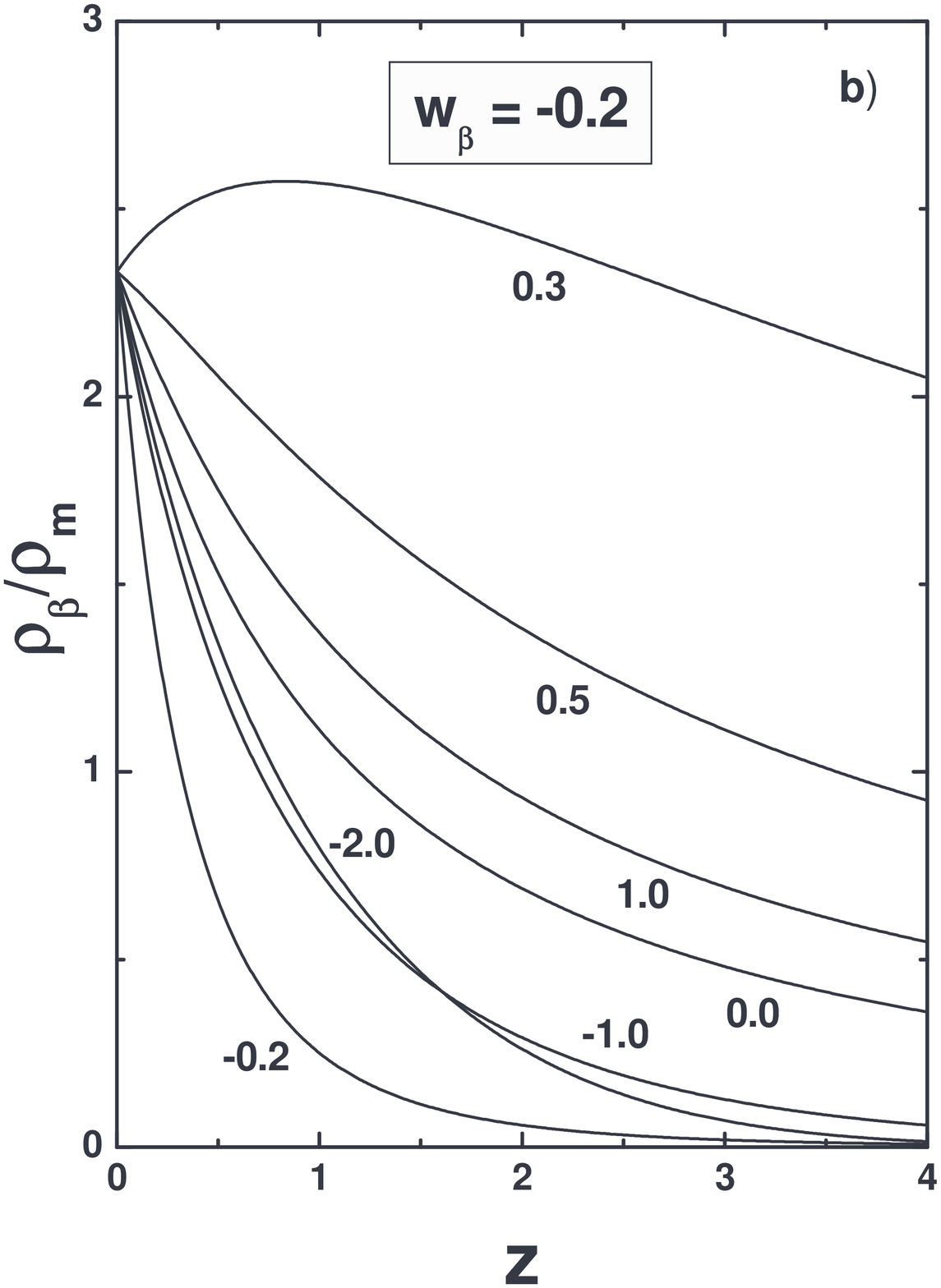,width=3.4truein,height=2.6truein,angle=0} 
\hskip 0.1in}
\caption{{\bf{a)}} The ratio $\rho_{\beta}/\rho_{m}$ as a function of the redshift parameter $z$ for $w_0 = -1.0$, $w_\beta = 0.2$ and $\rho_\beta^0/\rho_{m}^0 \simeq 2.33$. The value of $\beta$ is displayed below the corresponding curve. Note that the class of $\rm{P}_{\beta < 0}$ models presents an undesirable behaviour at high-$z$ in agreement with Eq. (\ref{DEDensity}). {\bf{b)}} The same as in Panel (a) for $w_\beta = - 0.2$. In this case, the dark energy contribution for negative values of $\beta$ becomes negligible at $z \gtrsim 4$.}
\end{figure*}

The Taylor expansion P1 was suggested in Refs. \cite{Huterer,astier,Weller}. Observational constraints on P1 were firstly studied in Ref. \cite{Huterer} by using SNe Ia data, gravitational lensing statistics and globular clusters ages and also in~\cite{goliah} that investigated limits to this parameterization from future SNe Ia experiments. As commented in \cite{Huterer}, P1 is a good approximation for most quintessence models out to redshift of a few and it is exact for models where the EoS is a constant or changing slowly. P1, however, has serious problems to explain high-$z$ observations since it  blows up at $z > 1$ as $\exp{(3w_{\rm{P1}}z)}$ for values of $w_{\rm{P1}} > 0$. The empirical fit P2 was introduced by Efstathiou~\cite{Efstathiou} who argued that for a wide class of potentials associated with dynamical scalar field models the evolution of $w(z)$ at $z \lesssim 4$ is well approximated by P2. P3 was proposed in Refs. \cite{Chevallier, Linder} aiming at solving undesirable behaviours of P1 at high-$z$. According to \cite{linde}, this parameterization is a good fit for many theoretically conceivable scalar field potentials, as well as for small recent deviations from a pure cosmological constant behaviour ($w= -1$).

To extend the EoS parametrizations above, let us consider the following time-dependent function:

\begin{eqnarray}
\label{Parametrization_a}
w(a)& = & w_0-w_{\beta}\frac{a^{\beta}-1}{\beta}\quad \quad \quad \quad (\rm{P_\beta}) \nonumber \\
& = &w_0-w_{\beta}\frac{(1+z)^{-\beta}-1}{\beta} \;,
\end{eqnarray}
where $a =  1/(1+z)$ is the cosmological scale factor and we have set its present value $a_0 = 1$ (throughout this paper both subscript and superscript 0 will denote present values). From the above expressions, it is straightforward to show that the EoS parameterizations given by Eq. (\ref{ps}) are fully recovered in the limits:
\begin{eqnarray}
\label{Ps}
\begin{tabular}{l}
$\beta \rightarrow -1$ \quad  \quad $\Rightarrow$  \quad  \quad   \hspace{0.17cm} $\rm{P_\beta}  \rightarrow P1$\;,\\
\\
$\beta \rightarrow 0$ \quad  \quad \quad $\Rightarrow$  \quad   \hspace{0.43cm} $\rm{P_\beta}  \rightarrow P2$\;,\\
\\
$\beta \rightarrow +1$ \quad  \quad  $\Rightarrow$ \quad  \quad   \hspace{0.17cm} $\rm{P_\beta}  \rightarrow P3$\;,
\end{tabular}
\nonumber
\end{eqnarray}
where we have used the equality $\ln{x} = \lim_{\xi \rightarrow 0} (x^{\xi} - 1)/\xi$ to obtain the limit for P2. This amounts to saying that the introduction of the new parameter $\beta$ is equivalent to insert the EoS parameterizations (P1)-(P3) in a more general framework that admits a wider and new range of cosmological solutions (${\rm{P}}_{\beta \gtrless 0}$).

Since the above parameterizations represent separately conserved components, one can show from the energy conservation law [$\dot{\rho}_\beta = -3{\dot{a}}(\rho_\beta+p_\beta)/a$]  that the ratio $f_\beta=\rho_\beta/\rho_{\beta}^0$ evolves as
\begin{equation}
\label{DEDensity}
f_\beta=a^{-3(1+w_0+w_{\beta}/\beta)}\exp\Big[\frac{3w_{\beta}}{\beta}\big(\frac{a^{\beta}-1}{\beta}\big) \Big]\;.
\end{equation}
Some special cases of the above expression are:
\begin{subequations} \label{101}
\begin{eqnarray} 
f_{\rm{P1}} = a^{-3(1+w_0-w_1)}\exp\Big[3w_1\big(\frac{1}{a}-1\big) \Big],\,\,\mbox{if}\,\,\,\beta=-1, \\
\nonumber \\
f_{\rm{P2}} = a^{-3[1+w_0-(w_2/2)\ln a]},\,\,\mbox{if}\,\,\,\beta=0, \\
\nonumber  \\
f_{\rm{P3}} =a^{-3(1+w_0+w_3)}\exp\Big[3w_3\big(a-1\big) \Big],\,\,\mbox{if}\,\,\,\beta=1. 
\end{eqnarray}
\end{subequations}

From Eq. (\ref{DEDensity}), some cases of interest relating the parameters $w_0$, $w_{\beta}$ and $\beta$ may be obtained:

\begin{enumerate}
 
\item $\beta > 0$ ($w_{\beta} \gtrless 0$): At early times the dark energy is a subdominant component if $w_0 + w_{\beta}/\beta \leq 0$.

\item $\beta < 0$ and $w_{\beta} > 0$: At early times the dark energy always dominates over the other material components.

\item $\beta < 0$ and $w_{\beta} < 0$: At early times the dark energy density vanishes.

\end{enumerate}

To better visualize the cases discussed above, we show in Figs. 1a and 1b the ratio $\rho_{\beta}/\rho_{m}$ as a function of $z$ for some selected values of $\beta$, $w_0 = -1$ and $\rho_{\beta}^0/\rho_{m}^0 \simeq 2.33$. Two symmetric values of $w_{\beta}$ are considered, i.e., 0.2 (Fig. 1a) and -0.2 (Fig. 1b) and the corresponding value of $\beta$ is displayed right below the curve. We observe that, for these particular combinations of $w_0$ and $w_{\beta}$, almost the entire range of $\rm{P}_{\beta \geq 0}$ solutions (which includes P2 and P3) are well-behaved~\footnote{Note that, although well-behaved in the past evolution, P3 blows up exponentially in the future as $z \rightarrow -1$ for $w_{\rm{P3}} > 0$. In general, for $\beta > 0$ and $w_{\beta} < 0$, $\rho_{\beta} \rightarrow 0$ as $z \rightarrow -1$.}. As expected, for $w_{\beta} = 0.2$ ($w_{\beta} > 0$ in general) $\rm{P}_{\beta < 0}$ parameterizations present undesirable behavior due to the exponential term in Eq. (4a). 

We also note that $\rm{P_\beta}$ is flexible enough to incorporate into it other dark energy scenarios. For example, models well approximated by P3 (or, equivalently, $\rm{P_{\beta = 1}}$) with $w_3 = {\rm{const.}}(1 + w_0)$ are clearly particular examples of $\rm{P_\beta}$. This is the case of the linear potential scenario of the type $V(\phi) = V_0 + (\phi - \phi_0)V'_0$ studied in Ref.~\cite{linear} and also of the so-called \emph{mirage} $\Lambda$ model of Ref.~\cite{mirage}. Still in the class of thawing scalar field models, the dynamics of the Pseudo-Nambu Goldstone Boson (PNGB) model~\cite{pngb}, whose potential is given by $V(\phi) \propto 1 + \cos{(\phi/f)}$, can be approximated by $w(a) = -1 + (1 + w_0)a^F$, where $F$ is inversely related to the symmetry scale $f$~\cite{rubin}. This three-parameter EoS can be incorporated into $\rm{P_\beta}$ by redefining $w_\beta = -\beta(1 + w_0)$. A similar identification can also be made to the class of thawing models studied in Ref.~\cite{jsa} whose potential is described by $V(\phi) \propto f(\phi) \exp[-\lambda(\phi + \alpha \phi^2)]$ and EoS given exactly by $w(a) = -1 + \lambda a^{2\alpha}$. Clearly, $\rm{P_\beta}$ can reproduce this latter $w(a)$ function by redefining $\lambda = w_\beta/\beta$ with the constraint $w_0 - w_\beta/\beta = -1$ (we refer the reader to Refs. \cite{rubin,lh} for a complete analysis of several models discussed here and others that may potentially be described by $\rm{P_\beta}$)\footnote{Note that most of the trivial (three-parameter) generalizations of P1 - P3 can be incorporated into $\rm{P_\beta}$. For example, let us take the case of the model $w(a) = w_0 + w_{\rm{P3}}(1 - a^b)$ discussed in Ref.~\cite{lh}, which is clearly a particular case of $\rm{P_\beta}$ when $w_\beta = \beta w_{\rm{P3}}$.}.

Finally, the Friedmann equation for our generalized $w_{\beta}(z)$ model is written as
\begin{eqnarray}
\label{FriedmannEquation_a}
{\cal{H}}(z;{\bf s})=\sqrt{\Omega_m^0a^{-3} + (1 - \Omega_m^0 - \Omega_k)f_\beta + \Omega_k^0a^{-2}} \;,
\end{eqnarray}
where ${\cal{H}}(z;{\bf s}) = H/H_0$, ${\mathbf{s}} \equiv (\Omega^{0}_{m}, w_0, w_{\beta}, \beta)$, and $H_0$, $\Omega_m^0$ and $\Omega_k^0$ are, respectively, the current values of the Hubble, matter and curvature density parameters.

\section{Observational Aspects}

\subsection{Transition redshift}

In order to study the influence of the parameter $\beta$ on the epoch of cosmic acceleration, we first derive the deceleration parameter,
\begin{equation}
q(a) = \frac{1}{2}\frac{{\Omega_m^0a^{-3}} + (1 - \Omega_m^0 - \Omega_k)(f_{\beta}'/a - 2f_\beta) }{\Omega_m^0a^{-3} + (1 - \Omega_m^0 - \Omega_k)f_\beta + \Omega_k^0a^{-2}  }\;,
\end{equation}
where $f_{\beta}'/a = 3(1+\omega_0+{\omega_\beta\over\beta}){f_\beta}-{3\omega_\beta a^{\beta}\over\beta}
f_\beta $.
The transition redshift $z_t$, at which the Universe switches from deceleration to acceleration, can be obtained from the following expression
\begin{equation}
\label{zx}
\Omega_m^0y^3 + g_{\beta} (1 - \Omega_m^0 - \Omega_k^0)y^{3(1 + w_0 + \frac{w_{\beta}}{\beta})} = 0\;,
\end{equation}
where $y = (1+z_t)$ and 
$$
g_{\beta} = [1 + 3w_0 + \frac{3w_{\beta}}{\beta}(1-y^{-\beta})]\exp\left[{\frac{3w_{\beta}}{\beta}}\left(\frac{y^{-\beta} - 1}{\beta}\right)\right].
$$
As one may easily check, for values of $w_0 = -1$ and $w_{\beta} =0$, Eq. (\ref{zx}) reduces to the well-known standard expression $y_{\rm{\Lambda CDM}} = [2(1-\Omega_m^0)/\Omega_m^0]^{1/3}$.

Figure 2 shows the transition redshift $z_t$ as a function of $\beta$ [Eq. (\ref{zx})] by assuming $w_0 = -1$ and $\Omega_m^0 = 0.3$. Four cases are shown: two in which $w_{\beta}$ takes positive values (0.5 and 1.0) and two in which $w_{\beta} < 0$ (-0.5 and -1.0).  Note that, the more negative the value of $\beta$ the lower (higher) the transition redshift for negative (positive) values of $w_{\beta}$. The horizontal lines stand for the interval $0.49 \leq z_t \leq 0.88$, which corresponds to $\pm 1\sigma$ of the value for $z_t$ given in Ref.~\cite{davis}.

\begin{figure}
\centerline{\psfig{figure=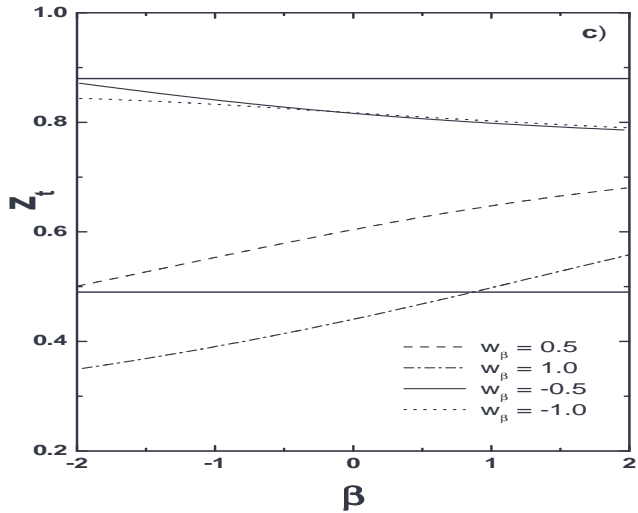,width=3.4truein,height=2.8truein,angle=0} 
\hskip 0.1in}
\caption{The influence of the parameter $\beta$ on the transition redshift $z_t$. To plot these curves we have fixed $w _0 = -1.0$ and $\Omega_{m}^0 = 0.3$. Solid horizontal lines stand for the interval $0.49 \leq z_t \leq 0.88$, which corresponds to $\pm 1\sigma$ of one of the values for $z_t$ estimated in Ref.~\cite{davis}.}
\end{figure}

\subsection{Statistical analysis}

The new parameter $\beta$ opens the possibility for a multitude of new cosmological solutions for different combinations of $w_0$, $w_{\beta}$ and $\beta$. In this Section we investigate observational bounds on the parametric spaces $w_0 - \beta$, $w_\beta - \beta$ and $w_0 - w_{\beta} - \beta$ from a statistical analysis involving four classes of cosmological observations.

Motivated by inflation and the recent results of the CMB power spectrum~\cite{wmap} we assume from now on spatial flatness ($\Omega_k = 0$). We use the most recent compilation of distance measurements to SNe Ia, the so-called {Constitution set} (CS)~\cite{cs} of 397 SNe Ia. This SNe Ia sample covers a redshift range from $z = 0.015$ to $z = 1.551$, including 139 SNe Ia at $z < 0.08$, and constitutes the largest SNe Ia luminosity distance sample currently available. 

We also use CMB and BAO data to help diminish the degeneracy between the dark energy parameters $w_0$, $w_{\beta}$ and $\beta$. For the CMB, we use only the measurement of the CMB shift parameter~\cite{wmap,wang}
\begin{equation}
\label{ShiftParameter}
\mathcal{R}\equiv\sqrt{\Omega_m^0}\int_0^{z_{\mathrm ls}}\frac{dz'}{{\cal{H}}(z';{\bf s})} = 1.70\pm 0.03,
\end{equation}
where $z_{\mathrm ls} = 1089$ is the redshift of the last  scattering surface. The BAO parameter is given by~\cite{bao} 
\begin{equation}
{\cal{A}} = D_V\frac{\sqrt{\Omega_{mo} H_0^2}}{z_*}\;,
\end{equation}
where the SDSS value is ${\cal{A}}_{obs} = 0.469 \pm 0.017$, $z_* = 0.35$ is the typical redshift of the SDSS sample and $D_V = [D_M^2/{z_*}{H(z_*;{\mathbf s})}]^{1/3}$ is the dilation scale, defined  in terms of the comoving distance to $z_*$, i.e., $D_M = \int_{0}^{z_*}{dz'}/{H(z';{\mathbf s})}$. It is worth emphasizing that the value of ${\cal{A}}$ is obtained from the data in the context of the $\Lambda$CDM model, and can be considered a good approximation only for models whose dark energy contribution at early times is not very large \cite{doran}. Therefore, since for values of $\beta < 0$ and $w_\beta > 0$ $\rm{P_\beta}$ gives rise to early dark energy models, we must have in mind that the inclusion of BAO data (as well as the CMB shift parameter) rigorously limits the range of the parameter space considered in the analysis.

Finally, we also use 9 determinations of the Hubble parameter as a function of redshift, as given in Ref. \citep{jimenez}. The use of these data to constrain cosmological models seems to be interesting because, differently from distance measures, the Hubble parameter is not integrated over (see, e.g.,~\citep{jimenez,hz} for more details). Thus, in our statistical analysis we minimize the function
$\chi^2 = \chi^{2}_{\rm{SNe}} + \chi^{2}_{\rm{CMB}} + \chi^{2}_{\rm{BAO}} + \chi^{2}_{\rm{H(z)}}$, which takes into account all the data sets discussed above.

\begin{figure}[t]
\centerline{
\psfig{figure=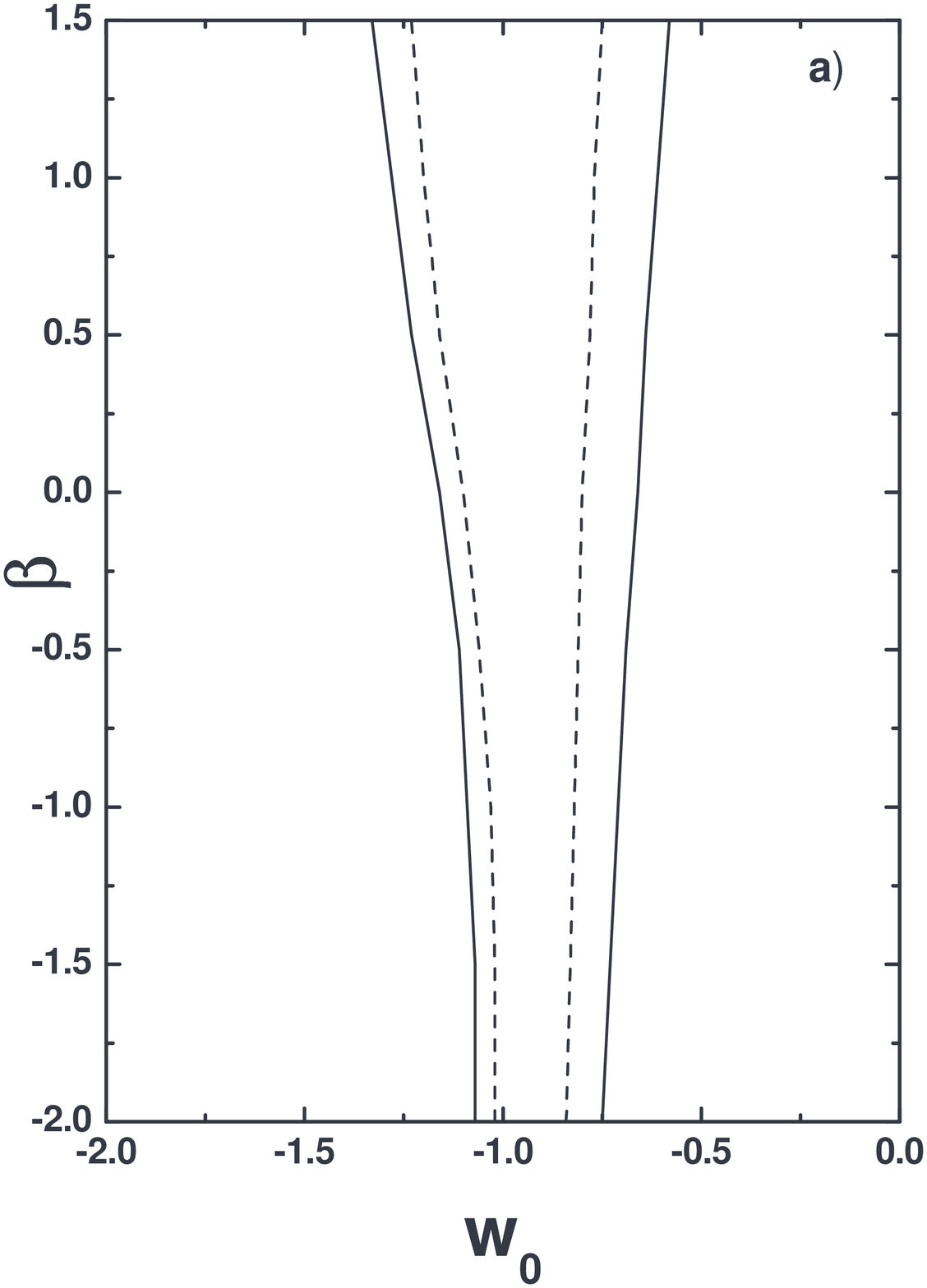,width=1.8truein,height=2.8truein,angle=0} 
\psfig{figure=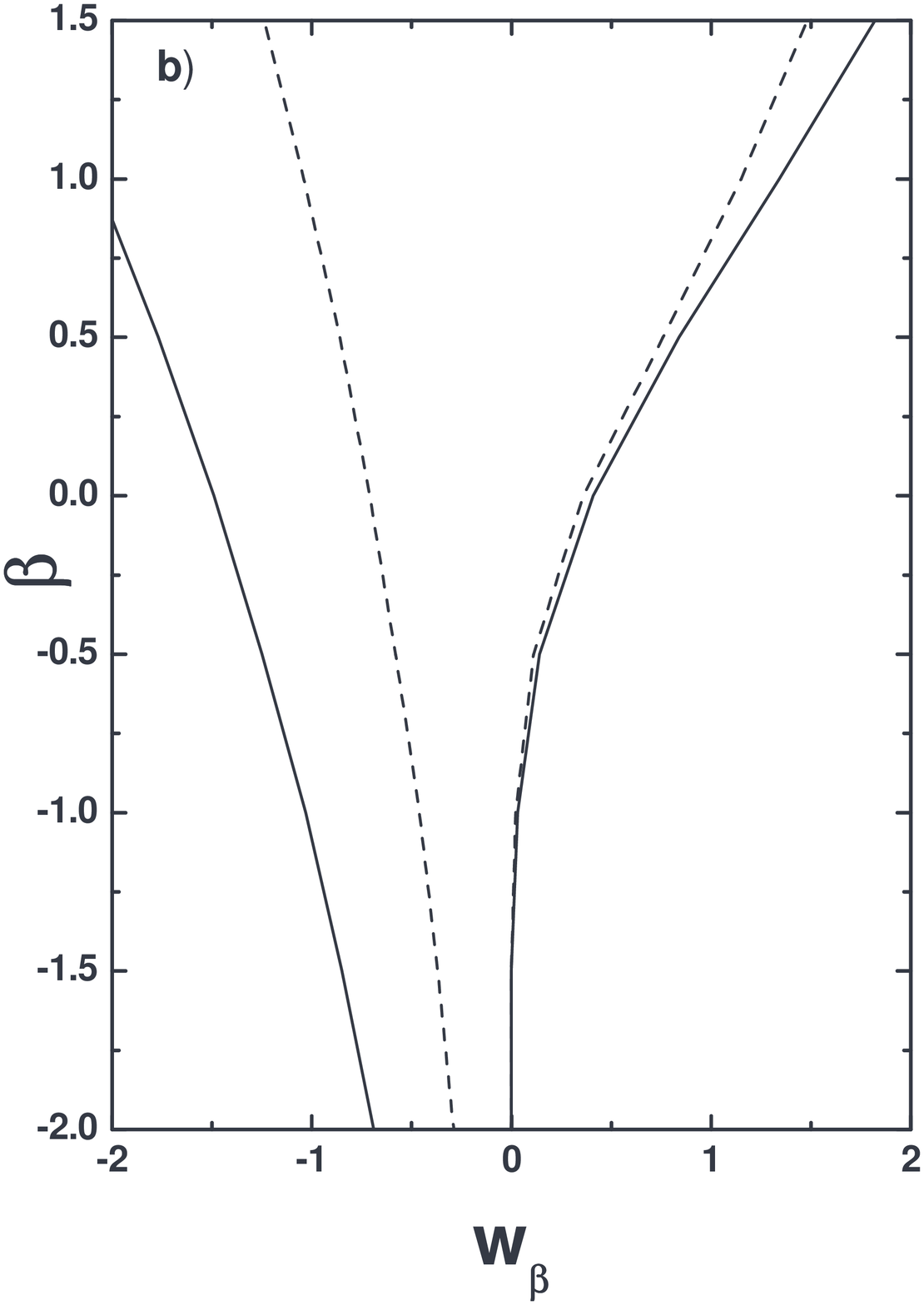,width=1.8truein,height=2.8truein,angle=0}
\hskip 0.1in}
\caption{Contours of $\chi^2$ in the plane $w_0 - \beta$ (Panel a) and $w_\beta - \beta$ (Panel b). Contours are drawn for $\Delta\chi^2 = 2.30$ (1$\sigma$) and 6.17 (2$\sigma$). From Panel (b), we clearly see that the observational data are compatible with the class of P$_{\beta < 0}$ models predominantly for values of $w_\beta < 0$. }
\end{figure}
\begin{figure}
\centerline{\psfig{figure=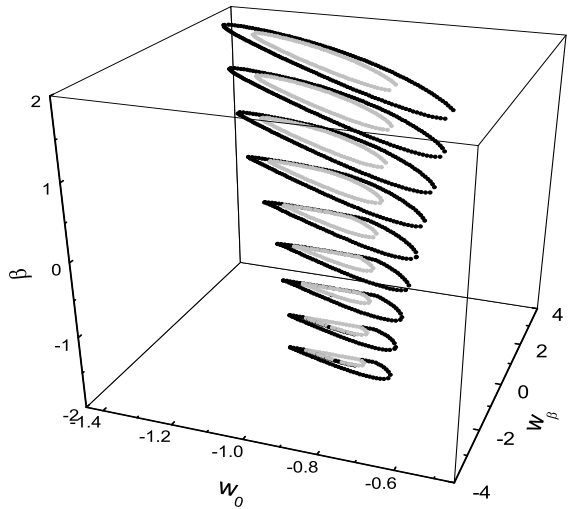,width=4.0truein,height=2.8truein,angle=0} 
\hskip 0.1in}
\caption{The 3-dimensional plane $w_0 - w_\beta - \beta$ from SNe Ia + BAO + CMB data. Contours are drawn for $\Delta \chi^2 = 3.53$ (1$\sigma$) and 8.02 (2$\sigma$).  The best-fit  occurs for values of $w_0 \simeq -1.0$, $w_{\beta} \simeq 0.28$ and $\beta \simeq 0.1$ with $\chi^2_{\nu} \simeq 1.17$.}
\end{figure}

\subsection{Results}

Figures 3 shows the parametric spaces $w_0 - \beta$ and $w_{\beta} - \beta$ that arise from the joint analysis described above. As expected, we note that similarly to what happens with most of the time-dependent EoS parameterizations the current observational bounds on $w_{\beta}$ and $\beta$ are quite weak since they appear as the argument of the exponential term in the energy density [Eq. (\ref{DEDensity})]. Due to the CMB shift estimate $\mathcal{R}$ at high-$z$, we see from Panel (3b) that the observational data are compatible with the class of $\rm{P}_{\beta < 0}$ models predominantly for values of $w_{\beta} < 0$, which is compatible with the cases of interest discussed in Sec. II and also with the $\rho_{\beta}/\rho_m$ -$z$ history shown in Fig. (1a)-(1b). We also show in Fig. 4 contours of $\Delta \chi^2$ in the 3-dimensional parametric space $w_0 - w_{\beta} - \beta$, using all data sets. The contours are drawn for $\Delta \chi^2 = 3.53$ and 8.02 (corresponding, respectively, to 1$\sigma$ and 2$\sigma$ for 3 parameters).  In particular, for the combination of data discussed earlier, the best-fit  occurs for values of $w_0 \simeq -1.0$, $w_{\beta} \simeq 0.28$ and $\beta \simeq 0.1$ with $\chi^2_{\nu} \simeq 1.17$ ($\chi^2_{\nu} \equiv \chi^2_{min}/\nu$ where $\nu$ stands for degrees of freedom). We note that, when the CMB shift parameter is not considered in the $\chi^2$ analysis [SNe Ia + BAO + H($z$)], the best-fit for $\beta$ changes considerably to $\beta \simeq -3.04$ ($w_0 \simeq -0.98$, $w_{\beta} \simeq 0.1$). This difference in the results with and without the CMB shift estimate $\mathcal{R}$ seems to be in agreement with a recent analysis for P3 discussed in Ref.~\cite{sahni}.

\section{Final Remarks}

In principle, to check the validity of a model or a theory, it is interesting by several reasons to insert it in a more general framework. This not only brings to light new sets of solutions but also may provide a more accurate consistency check to the original model. In this paper, a general framework for a class of EoS parameterization (P1)-(P3), quantified by a dimensionless parameter $\beta$, has been proposed and some of its cosmological consequences studied. As an interesting consequence, we have shown that between (and beyond) P2 ($\beta = 0$) and P3 ($\beta = 1$), there is a family of $\rm{P}_{\beta > 0}$ solutions, whose behaviour seems to be compatible with current observational data. Although a reasonably precise estimate for $\beta$ (as well as for $w_{\beta}$) cannot be extracted from current data, we believe that the next generation of dark energy experiments dedicated to this issue [mainly those measuring the expansion history from high-$z$ SNe Ia, baryon oscillations, and weak gravitational lensing distortion by foreground galaxies (see, e.g., \cite{future}) will probe cosmology with sufficient accuracy to decide which (if any) interval of the parameters $w_{\beta}$ and $\beta$ is preferable from observational viewpoint (see also \cite{lh} for a discussion on some EoS parameterizations and possible constraints on their parameters from future SNe Ia, CMB and weak lensing experiments).

\begin{acknowledgments}

EMBJr, JSA and RS acknowledge financial support from CNPq-Brazil. Z-H Zhu was supported by the NNSFC under the Distinguished Young Scholar Grant 10825313, the Key Project Grant 10533010, and by the Ministry of Science and Technology National Basic  Science Program (Project 973) under grant No. 2007CB8154.

\end{acknowledgments}


\begin{thebibliography}{30}


\bibitem{rev}  V.~Sahni and A.~A.~Starobinsky, Int.\ J.\ Mod.\ Phys.\  D {\bf 9}, 373 (2000); P. J. E. Peebles and B. Ratra Rev. Mod. Phys. {\bf{75}}, 559 (2003);  T. Padmanabhan, Phys. Rept. {\bf{380}}, 235 (2003); E. J. Copeland, M. Sami and S. Tsujikawa, Int. J. Mod. Phys. {\bf{D15}}, 1753 (2006); J. S. Alcaniz, Braz. J. Phys. {\bf{36}}, 1109 (2006); J.~A.~Frieman, AIP Conf.\ Proc.\  {\bf 1057}, 87 (2008). arXiv:0904.1832 [astro-ph.CO].

\bibitem{quint} R. R. Caldwell, R. Dave and P. J. Steinhardt, Phys. Rev. Lett. {\bf{80}}, 1582 (1998); P.G. Ferreira and M. Joyce, Phys. Rev. {\bf{D58}}, 023503 (1998); R.~R.~Caldwell and E.~V.~Linder, Phys.\ Rev.\ Lett.\  {\bf 95}, 141301 (2005). J.S. Alcaniz, R.Silva, F.C. Carvalho, Z.-H. Zhu, Class. Quantum Grav. {\bf{26}}, 105023 (2009). arXiv:0807.2633 [astro-ph]

\bibitem{phantom} R. R. Caldwell, Phys. Lett. B 545, 23 (2002); V. Faraoni, Int. J. Mod. Phys. D 11, 471 (2002); S. M. Carroll, M. Hoffman, and M. Trodden, Phys. Rev. D 68, 023509 (2003); J. S. Alcaniz, Phys. Rev. D 69, 083521 (2004); S. Nesseris and L. Perivolaropoulos, Phys. Rev. D 70, 123529 (2004); J. Santos and J. S. Alcaniz, Phys. Lett. B 619, 11 (2005).

\bibitem{ec} S. W. Hawking and G. F. R. Ellis, The Large Scale Structure of Spacetime (Cambridge University Press, Cambridge, England, 1973).


\bibitem{teg} Y. Wang and M. Tegmark, Phys. Rev. Lett. {\bf{92}}, 241302 (2004); E.~M.~Barboza and J.~S.~Alcaniz, Phys.\ Lett.\  B {\bf 666}, 415 (2008).

\bibitem{Huterer} A.~R.~Cooray and D.~Huterer, Astrophys.\ J.\  {\bf 513}, L95 (1999).

\bibitem{astier} P.~Astier, arXiv:astro-ph/0008306.

\bibitem{Weller} Weller, J. and Albrecht, A., Phys. Rev Lett. {\bf 86}, 1939 (2001).


\bibitem{Efstathiou} G. Efstathiou, MNRAS {\bf 342}, 801 (2000).

\bibitem{Chevallier} M. Chavallier and D. Polarski, Int. J. Mod. Phys. D {\bf 10}, 213 (2001).

\bibitem{Linder}  E. V. Linder, Phys. Rev Lett. {\bf 90}, 091301 (2003).

\bibitem{goliah} M. Goliath {\it et al.}, Astron. Astrophys. {\bf{380}}, 6. (2001).

\bibitem{otherP} Y. Wang and P.M. Garnavich, Astrophys. J. {\bf{552}}, 445 (2001); C. R. Watson and R. J. Scherrer, Phys. Rev. D {\bf{68}}, 123524 (2003); P.S. Corasaniti et al., Phys. Rev. D {\bf{70}}, 083006 (2004); H.K. Jassal, J.S. Bagla and T. Padmanabhan, Mon. Not. R. Astron. Soc. {\bf{356}}, L11 (2005); V. B. Johri and P. K. Rath, Phys. Rev. D {\bf{74}}, 123516 (2006).

\bibitem{l1} E.~V.~Linder, Rept.\ Prog.\ Phys.\  {\bf 71}, 056901 (2008).



\bibitem{linde} J. Kratochvil {\it et al.}, JCAP {\bf{0407}}, 001 (2004).

\bibitem{linear} A. Linde in ``Three Hundred Years of Gravitation'', Eds. S. Hawking and W. Israel (Cambridge: Cambridge Univ. Press.), 604 (1987); S. Weinberg, Cosmology (Oxford: Oxford Univ. Press.), 2008.

\bibitem{mirage} E.~V.~Linder,  ``The Mirage of w=-1,''  arXiv:0708.0024 [astro-ph].

\bibitem{pngb} J.~A.~Frieman, C.~T.~Hill, A.~Stebbins and I.~Waga, Phys.\ Rev.\ Lett.\  {\bf 75}, 2077 (1995).

\bibitem{rubin} D.~Rubin {\it et al.}, Astrophys.\ J.\  {\bf 695}, 391 (2009).

\bibitem{jsa}  F.~C.~Carvalho, J.~S.~Alcaniz, J.~A.~S.~Lima and R.~Silva, Phys.\ Rev.\ Lett.\  {\bf 97}, 081301 (2006).

\bibitem{lh}  E.~V.~Linder and D.~Huterer, Phys.\ Rev.\  D {\bf 72}, 043509 (2005).


\bibitem{davis}  J.~V.~Cunha,
  Phys.\ Rev.\  D {\bf 79}, 047301 (2009). 


\bibitem{wmap} D. N. Spergel {\it et al.}, Astrop. J. Suppl. {\bf{148}}, 175 (2003); D. N. Spergel {\it et al.}, \apj Supl. {\bf 170}, 377 (2007); J.~Dunkley {\it et al.}, arXiv:0803.0586 [astro-ph].

\bibitem{cs} M.~Hicken {\it et al.}, ``Improved Dark Energy Constraints from ~100 New CfA Supernova Type Ia Light Curves,''. 
arXiv:0901.4804 [astro-ph.CO]


\bibitem{wang} Y. Wang, and P. Mukherjee, Astrophys. J. {\bf 650}, 1 (2006).


\bibitem{bao} D.J. Eisenstein {\it et al.}, Astrophys. J. {\bf{633}}, 560 (2005).

\bibitem{doran} M. Doran, S. Stern, and E. Thommes, J. Cosmol. Astropart. Phys. {\bf{04}} (2007) 015.

\bibitem{jimenez} J.~Simon, L.~Verde and R.~Jimenez,  Phys.\ Rev.\  D {\bf 71}, 123001 (2005).


\bibitem{hz} L.~Samushia and B.~Ratra, Astrophys.\ J.\  {\bf 650}, L5 (2006); Z.~L.~Yi and T.~J.~Zhang, Mod.\ Phys.\ Lett.\  A {\bf 22}, 41 (2007); F.~C.~Carvalho {\it et al.}, JCAP {\bf 0809}, 008 (2008); H.~Zhang, H.~Yu, H.~Noh and Z.~H.~Zhu,  Phys.\ Lett.\  B {\bf 665}, 319 (2008); M.~A.~Dantas and J.~S.~Alcaniz, arXiv:0901.2327 [astro-ph.CO]; F.~E.~M.~Costa, E.~M.~Barboza and J.~S.~Alcaniz, arXiv:0905.0672 [astro-ph.CO].



\bibitem{future} A. Crotts et al., astro-ph/0507043; see also http://jedi.nhn.ou.edu/.

\bibitem{sahni} A.~Shafieloo, V.~Sahni and A.~A.~Starobinsky,
  arXiv:0903.5141 [astro-ph.CO].


\end{thebibliography}
\end{document}